\documentclass[preprint]{aastex}
\usepackage{graphicx}

\usepackage{natbib}

\newcommand{\be}{\begin{equation}}
\newcommand{\ee}{\end{equation}}
\newcommand{\nn}{\mbox{} \nonumber \\ \mbox{} }
\newcommand{\ba}{\begin{eqnarray}}
\newcommand{\ea}{\end{eqnarray}}

\newcommand{\Bf}{{magnetic field}}

\newcommand\eg{{\it{e.g.}}}

\newcommand\lo{\mathrel{\raise.3ex\hbox{$<$}\mkern-14mu\lower0.6ex\hbox{$\sim$}}}
\newcommand\go{\mathrel{\raise.3ex\hbox{$>$}\mkern-14mu\lower0.6ex\hbox{$\sim$}}}

\begin{document}
%\runauthor{Lyutikov}
%\begin{frontmatter}
\title{Rotating Parker wind
}

\author{Maxim Lyutikov}

\affil{
Department of Physics and Astronomy, Purdue University, 
 525 Northwestern Avenue,
 West Lafayette, IN
47907-2036 }

\begin{abstract}
We reconsider the structure of thermally driven rotating   Parker wind.   Rotation, without \Bf, changes qualitatively the structure of the subsonic region: solutions become non-monotonic and  do not extend to the origin.  For small  angular velocities solutions have two critical  points -  X-point and  O-points, which merge  at the  
 critical angular velocity of the central star 
$\Omega_{crit} = G M_\ast/(2 \sqrt{2} c_s R_{b}^2)$ (where $M_\ast$ and $R_{b}$ are mass and radius of the central star, $c_s$ is the sound speed in the wind). For larger spins there is no critical points in the solution.
For disk winds (when the base of the wind rotates with Keplerian velocity)  launched equatorially the coronal sound speed  should be smaller than $\approx 0.22 v_K$  in order to connect to the critical curve ($v_K$ is the  Keplerian velocity at a given location on the disk). 
\end{abstract}

\section{Introduction}

The Parker model of Solar wind \citep{1965SSRv....4..666P,1952MNRAS.112..195B} as well as  its MHD extension \citep{WeberDavis} are at the core of the stellar wind theory \citep{1999isw..book.....L}.  Of particular interests to us here  are thermally launched winds from rotating central objects -  a star or a disk.

This is a classical topic in stellar wind theory, that has not been considered to the best of our knowledge. 
Previously, a number of large scale 2D models of thermally-driven winds were constructed \citep[\eg][]{1990PASJ...42..249F,2016MNRAS.460.3044C,2012MNRAS.426.2239W}, but the basic equatorial flow of rotating thermally driven winds has not been properly  considered.
% \cite{2003ASSL..294.....G} \citep[see also][]{1999A&A...343..251K}  provided incorrect interpretation of the results, that there are two critical points in flow. In fact, there is   one or none critical points, as we demonstrate below. 
\cite{2010ApJS..188..290S} calculated numerically the structure of the rotating Parker winds  in {\it cylindrical geometry} -  our analytical results are in qualitative agreement with their work (see Appendix \ref{cyll} for comparing spherical and cylindrical outflows). Our analytical results also provide quantitative estimates for various wind regimes. 

%The generalization of the  Parker isothermal wind model to rotating unmagnetized case has not been considered to the best of our knowledge.  This is the purpose of the paper.

\section{Rotating Parker wind}

Consider a rotating star that launches thermally driven wind from its surface. The surface of a star may not have a clear physical definition - for mathematical purpose we define a surface at $R_{b}$ as  the base of the wind, which rotates with  a given angular velocity $\Omega$. 
In the  frame rotating with the star, in the equatorial plane,  and assuming axially symmetric flow, the governing equations are the Euler equation
\ba && 
v_r  v_r' - 2 \Omega v_\phi - \frac{v_\phi^2}{r}=r \Omega^2 +  \Phi'(r) - c_s^2 \partial _r \ln \rho
\nn &&
2 \Omega r + ( v_\phi r)'=0
\ea
and mass conservation 
\be
r^2 \rho v_r={\rm Constant}
\label{mdot}
\ee
Above $\Phi=   G M _\ast/r$ is the gravitational potential and other notations are standard. We assumed isothermal equation of state - polytropic equations of state introduce only mild modifications to the structure of the solutions for most polytropic indices of interest  \citep[\eg][]{1999isw..book.....L}.

In the rotating frame, on the surface of the star $ v_\phi(R_{b})=0$, hence 
\be
v_\phi = \frac{\Omega R_{b}^2}{r} - r \Omega
\ee
(In the observer frame only the  first term remains, the main equation (\ref{1}) remains unchanged.)

The radial component becomes
\be
\frac{v_r '}{v_r}  = \frac{ 2 c_s^2 r^2 + r^3 \Phi' (r) + R_{b}^4 \Omega^2}{r^3(v_r^2 -c_s^2)}
\label{1}
\ee
This is the generalization of Parker-Bondi  equation, it is the main equation to be studied in the present paper. It is of critical point-type behavior. It differs from the classical Parker-Bondi case by the term $\Omega ^2$ in the numerator. With somewhat different notations, it agrees with \cite{2003ASSL..294.....G}, Eq. (6.44), see also \cite{1986ApJ...311..701F}.

\begin{figure}[h!]
\includegraphics[width=.99\columnwidth]{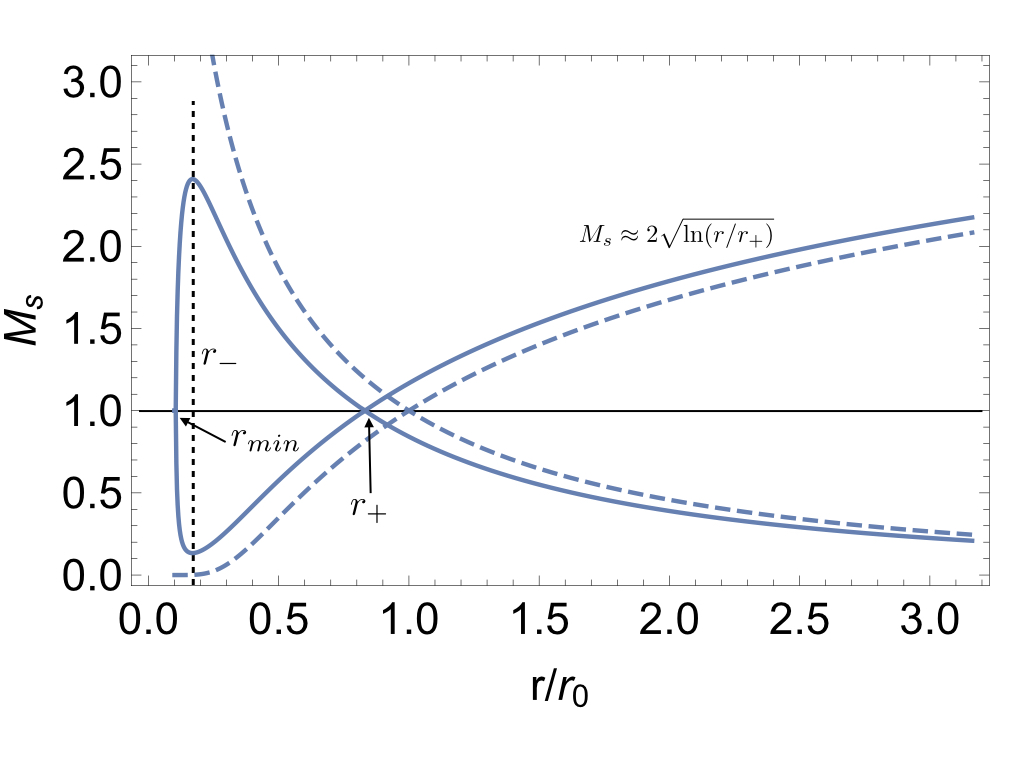} 
\caption{Phase diagram for Parker wind with rotation. Plotted is  sonic Mach number $M_s$ versus radius normalized to Parker-Bondi radius $r_0$; parameter  $R_\Omega = 0.75 r_0$. Dashed line is the conventional isothermal Parker wind.  Most important modifications in the rotating case is that the solution cannot extend to $r \rightarrow 0$. The large-$r$ asymptotic is not affected much. The value of $r_{min}$ is plotted in Fig. \protect\ref{rmin}; the Mach numbers at $r_{min}$ are plotted in Fig. \protect\ref{Msminmax}, evolution of density is plotted in Fig. \protect\ref{density}.}
 \label{Parker-rot}
\end{figure}

\begin{figure}[h!]
\includegraphics[width=.99\columnwidth]{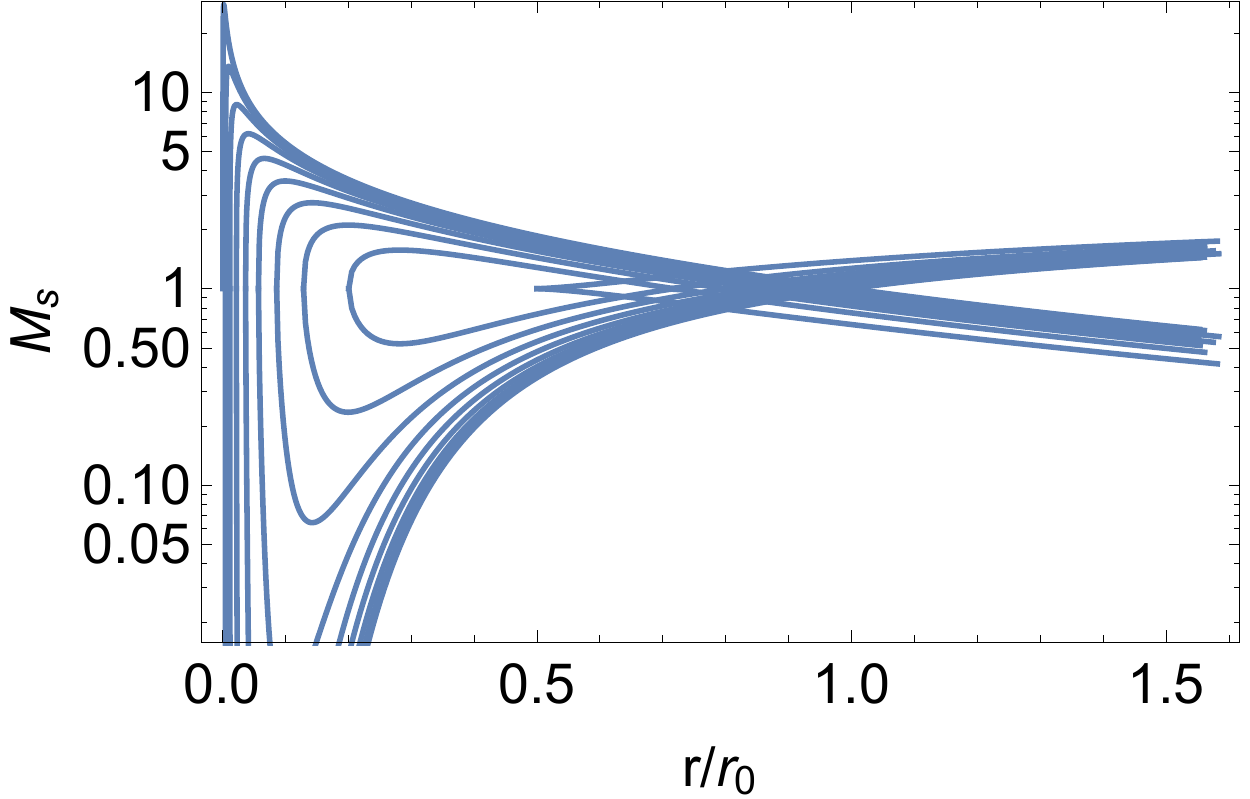} 
\caption{Same as Fig. \protect \ref{Parker-rot} but for different $R_\Omega= 0.1,0.2...1$.}
 \label{differentROmega}
\end{figure}

Introducing radial Mach number $M_s = v_r /c_s$,  Parker-Bondi radius  $r_0 =  G M_\ast/( 2 c_s^2)$
 and $R_\Omega = \sqrt{2} R_{b}^2 \Omega/c_s$, Eq. (\ref{1}) becomes
\be
(1-M_s^2) {\partial_r \ln M_s} = - \frac{ 2 r (r-r_0) +R_\Omega^2/2}{r^3}
\label{main}
\ee
There are two special points where both sides of Eq. (\ref{main}) are zero - $M_s=1$ 
and 
\be
r_{\pm } = \frac{1}{2} \left(r_0 \pm \sqrt{ r_0^2 - R_\Omega^2} \right) 
\label{rcrit}
\ee
(tidy form of these relations motivated our choice of normalization of $R_\Omega$).
The minus sign in (\ref{rcrit}) corresponds to the  O-type critical point, which has no implications for the dynamics of the wind.
The plus sign in (\ref{rcrit}) corresponds to the  X-type critical point, the  sub-to-supersonic transition, see Fig. \ref{Parker-rot} and Fig. \ref{differentROmega}.

General integral of (\ref{main}) is (Bernoulli function)
\be
B= 
\frac{M_s^2}{2}-\ln \left(M_s\right)-\frac{1}{2} \left(-\frac{R_{\Omega }^2}{2 r^2}+\frac{4 r_0}{r}+4 \ln
   (r)\right)
   \label{B}
   \ee
   The differential of the Bernoulli function vanishes only at $r_+$ - the only critical point in the flow.

The critical curves are given by
\be
-\frac{M_s^2}{2}  + \ln M_s = -\frac{1}{2}  +\left(  \frac{1}{ r^2} - \frac{1}{ r_+^2}\right) \frac{R_\Omega^2}{4} + 2   \left( \frac{1}{r_+}  -  \frac{1}{r}\right) r_0 +2 \ln \left( \frac{r_+}{r} \right)
\ee
Near the critical point $M_s =1,  \, r=r_+$, 
\be
M_s = 1 \pm \sqrt { 2- \frac{r_0}{r_+}} \left(1- \frac{r}{r_+} \right)
\ee
(the point $r_0 = 2 r_+$ is when the critical point disappears).

At $r \gg r_+$ the Mach numbers evolve according to $M\approx 2 \sqrt{\ln (r/r_+) }$ for supersonic branch and $M\approx (r/r_+)^{-2} $ for the subsonic branch, similar to the classical case of Parker wind.

\begin{figure}[h!]
\includegraphics[width=.99\columnwidth]{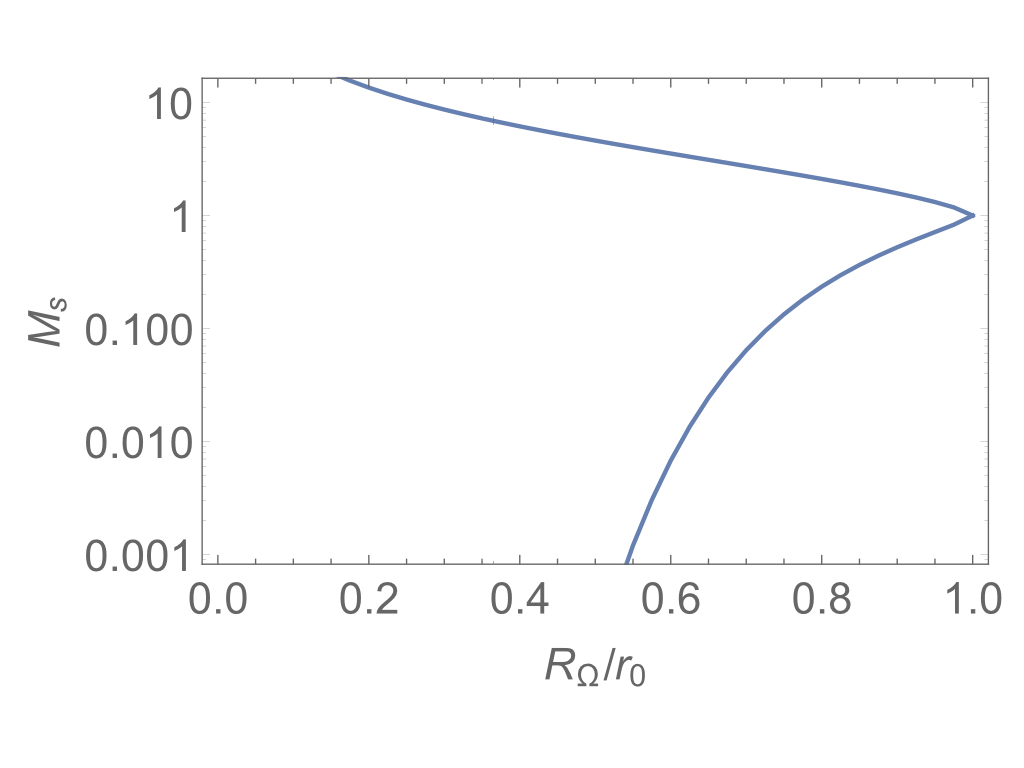} 
\caption{ Maximal and minimal values of Mach number in the  region $r \leq r_+$  along the critical curve as function of $R_\Omega$. }
 \label{Msminmax}
\end{figure}

The second point where the right hand side of  Eq. (\ref{main}) vanishes corresponds to points $r_-$, where $\partial_r M_s=0$. 
At the points $r_-$ the Mach number $M_s \neq 1$, see Fig.  \ref{Msminmax}.

\begin{figure}[h!]
\includegraphics[width=.99\columnwidth]{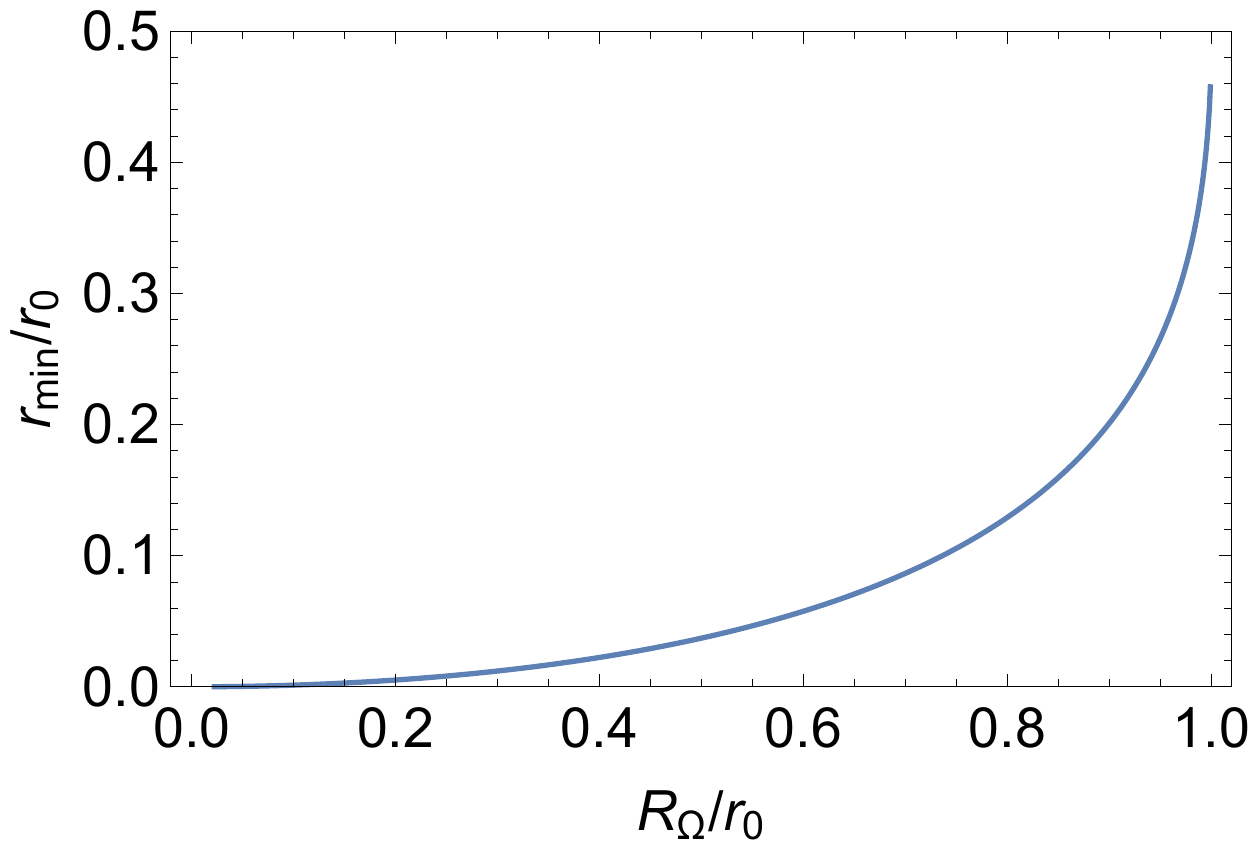} 
\caption{ Value of  $r_{min}$ for which the model is applicable as a function of $R_\Omega$.  For $R_\Omega=0$ we have $r_{min}=0$,  while for $R_\Omega=r_0$, $r_{min}=r_0/2$ and coincides at this moment with $r_+$.}
 \label{rmin}
\end{figure}

 Finally, 
setting $M_s=1$ gives two roots: one is $r_+$ - the critical point of the flow; another defines the minimal $r_{min}$ for which the model is applicable
\be
-\frac{R_{\Omega }^2}{4 r_{\min }^2}-2 \ln \left( \left(1+\sqrt{1-{R_{\Omega }^2}/{r_0^2}}\right)\frac{r_0}{2 r_{\min
   }}\right)+\frac{2 r_0}{r_{\min }}+\frac{R_{\Omega }^2}{r_0^2 \left(1+\sqrt{1-\frac{R_{\Omega
   }^2}{r_0^2}}\right){}^2}-\frac{4}{1+\sqrt{1-\frac{R_{\Omega }^2}{r_0^2}}}=0,
   \ee
 see Fig. \ref{rmin}. At $r_{min}$ we have $\partial_r M_s=\infty$. So, neither $r_-$ or $r_{min}$ are critical points (the phase curve is smooth and non-self-intersecting at those points).

For each given $R_\Omega$  the critical curve reaches at $r_-$  some maximal and minimal Mach numbers $M_{s, min/max}$, plotted in  Fig. \ref{Msminmax}. Mach number at $r_-$ are not unity, except  in the case $R_\Omega = r_0$, when points $r_\pm$ coincide.

\begin{figure}[h!]
\includegraphics[width=.49\columnwidth]{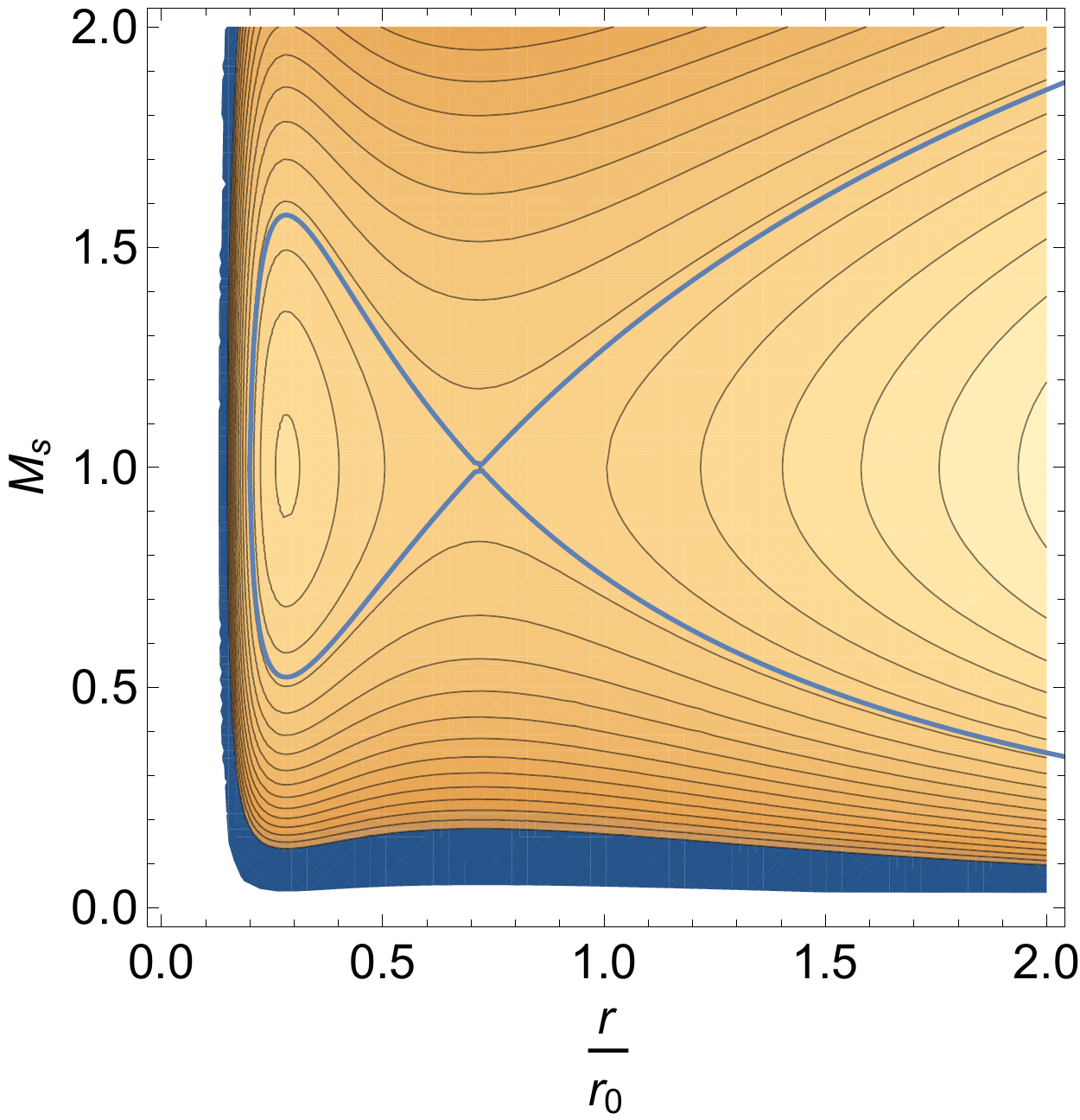} 
\includegraphics[width=.49\columnwidth]{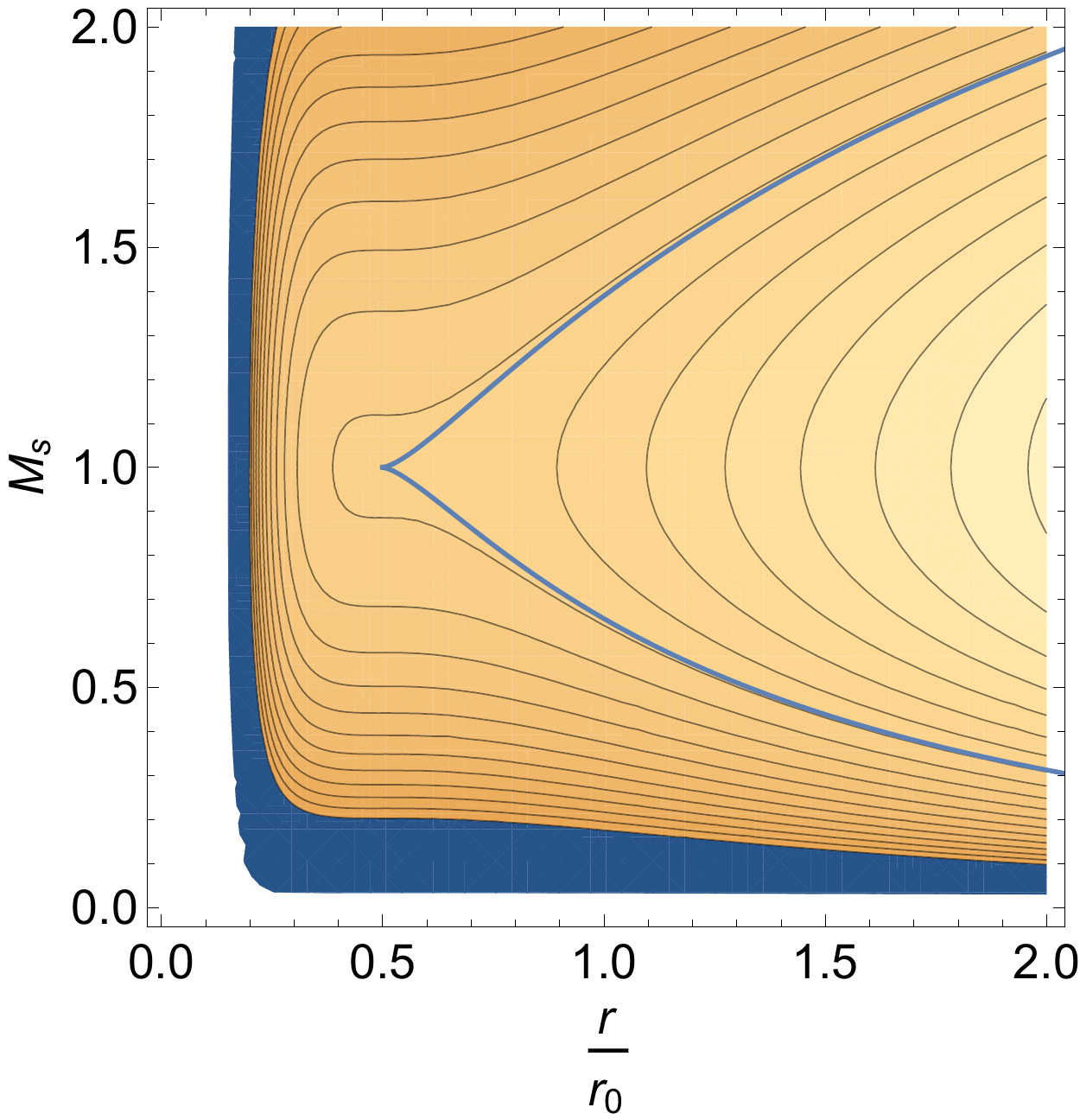} 
\caption{Phase curves on $r-M_s$ diagram for $R_\Omega =  0.9 r_0$ (left panel) and   $R_\Omega =   r_0$ (right panel).  Critical curves are highlighted.}
 \label{phase}
\end{figure}

Overall phase diagrams are plotted in Fig. \ref{phase}. For any  $0< R_\Omega< r_0$ there are  closed phase curves confined to  $r_{min}< r< r_+$.

\begin{figure}[h!]
\includegraphics[width=.49\columnwidth]{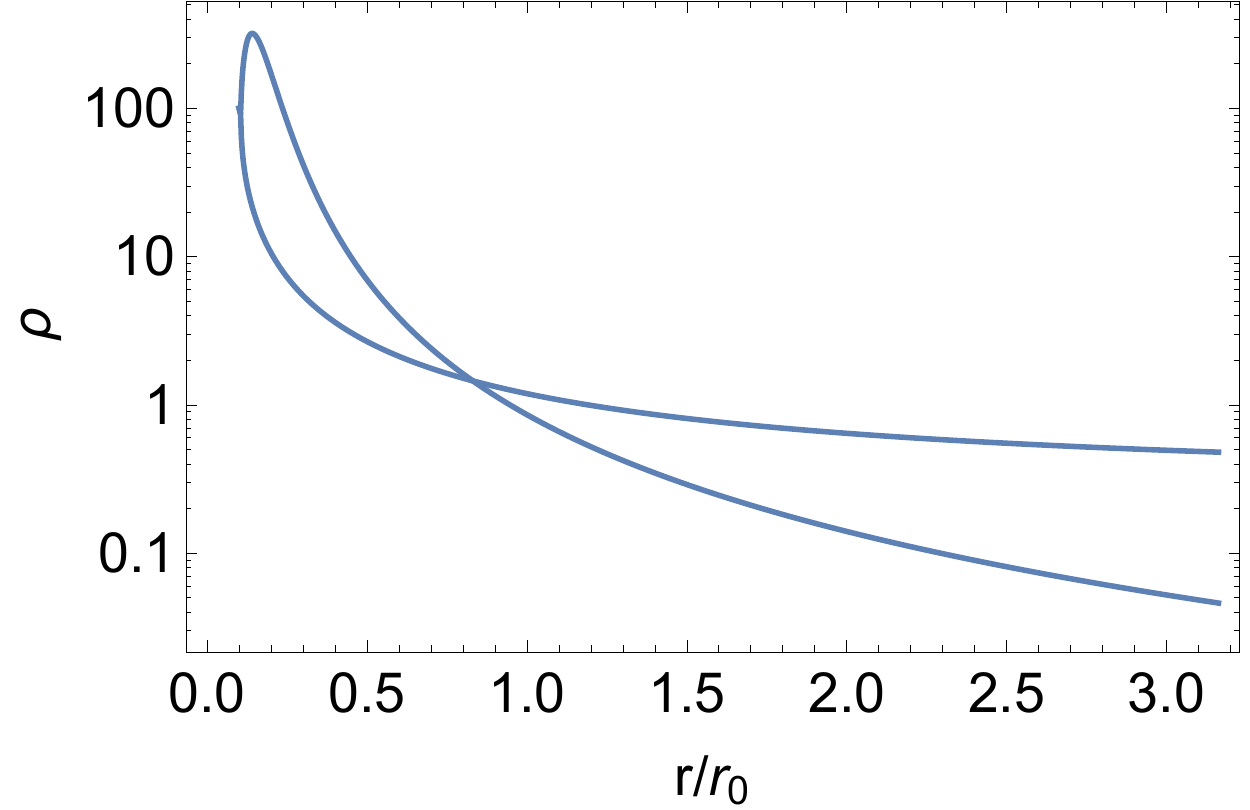} 
\includegraphics[width=.49\columnwidth]{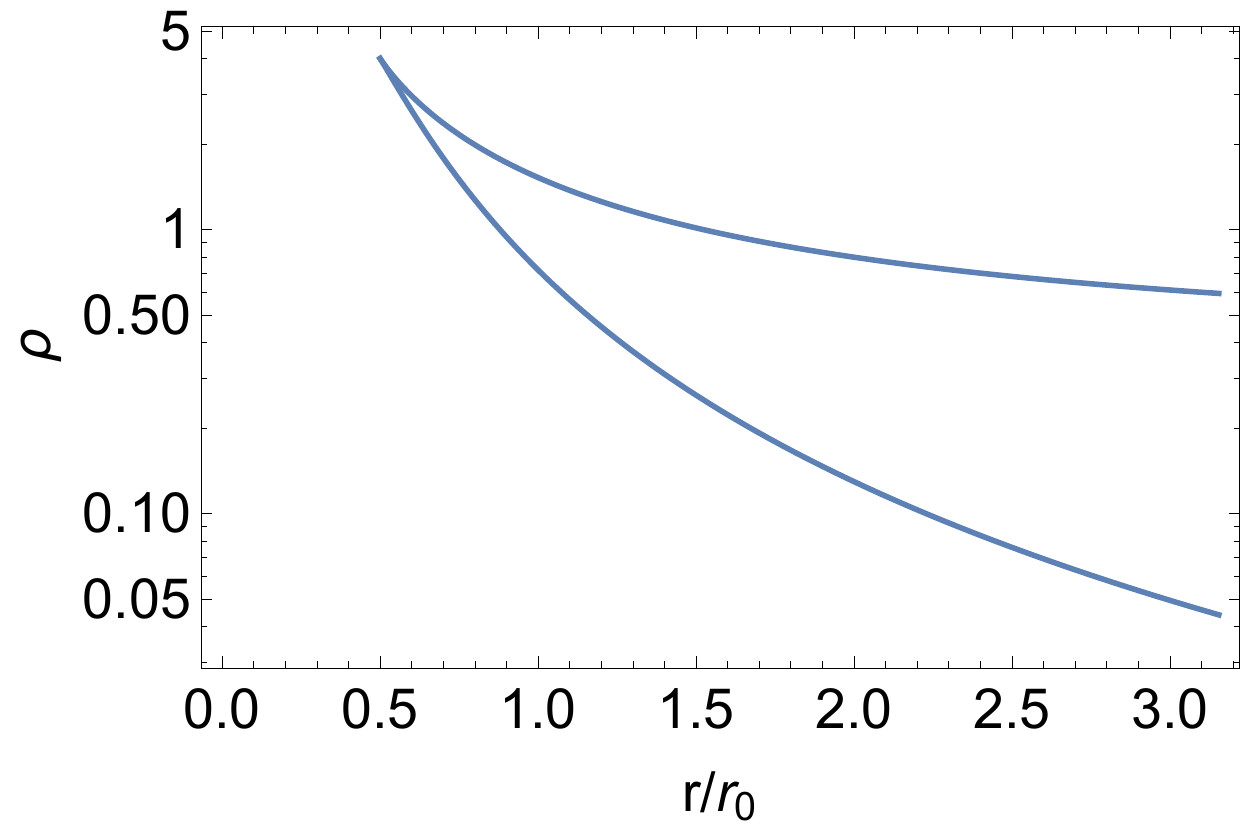} 
\caption{Density along the critical curves (plotted is the value of   $r_0^2/(M_s r^2)$) for $R_\Omega=0.75 r_0$ and  $R_\Omega= r_0$. At large radii larger densities correspond to the subsonic branch, where density reached a constant (since in that regime $M_s \propto r^{-2}$).}
 \label{density}
\end{figure}

The density along the critical curves of Fig. \ref{Parker-rot} are plotted in Fig. \ref{density}.

Given the velocity and density one can calculate the mass loss rate. It cannot be compared simply to the case of non-rotating winds. Usually, mass loss rate is calculated for a given base radius $R_b$ and local density - then the critical curve fixes the velocity. For rotating case, first,  there is a limit on the radius $R_b>r_{min}$, so such a procedure may not work, and, second, the similar procedure will give some different launching velocity at the same radius. 

\section{Constraints on the parameters}

\subsection{Existence of  the critical point}
There are several constraints on the parameters.  
Let's introduce two dimensionless parameters
\ba && 
\eta_\Omega= \frac{\Omega}{\Omega_K} \leq 1
\nn &&
\eta_s = \frac{c_s}{R_{b} \Omega_K}
\ea
where $\Omega_K$ is the break-up frequency at the equator.
We find
\ba &&
\frac{R_\Omega}{r_0} = 2 \sqrt{2} \eta _\Omega \eta_s
\nn &&
c_s = \frac{\eta_s}{\sqrt{2}} v_K
\ea
where $v_K = \sqrt{2 G M /r}$.

For the  points $r_\pm$ to exist, it is required that $R_\Omega < r_0$, which translates to 
\ba&&
\Omega \leq  \Omega_{crit,1} =\frac{1}{2 \sqrt{2}} \frac{GM}{ c_s R_{b}^2} = \frac{1}{2 \sqrt{2}}  \frac{R_{b} \Omega_K^2}{c_s}= \frac{1}{2 \sqrt{2} \eta_s} \Omega_K
\nn &&
 2 \sqrt{2} \eta _\Omega \eta_s\leq 1
 \label{const2}
\ea

Thus, there is a range  $1/(2 \sqrt{2}) < \eta_s < 1$ where $\Omega_{crit}< \Omega_{K}$. 
In physical units, this requires 
\be
c_s \geq \frac{1}{2 \sqrt{2}} \sqrt{ \frac{GM}{  R_{b}}}
\label{cs}
\ee
This is four times less that  the escape velocity without rotation $\sqrt{2 G M / R_{b}}$. Value (\ref{cs}) corresponds to the hydrodynamic energy parameter (HEP)  of \cite{2012MNRAS.426.2239W} $\lambda_0 = G M /(R_b c_s^2)= 8$. 
For larger   sound speed there is no critical point  and the flow always remains supersonic or subsonic depending on the conditions at the launch location $R_b$.

For  $R_\Omega=r_0$ we have
\ba &&
(1-M_s^2) \frac{\partial_r M_s}{M_s} = - \frac{  (2r-r_0)^2 }{2r^3}
\nn &&
\ln \left(M_s\right)-\frac{M_s^2}{2}=\frac{5}{2}+\frac{r_0^2}{4 r_{\min }^2}-\frac{2 r_0}{r_{\min }}+2 \ln
   \left(\frac{r_0}{2 r_{\min }}\right)
\ea
Thus, the numerator does not change sign - the outflow must start sonic at the critical  point $r=r_0/2$. For larger $R_\Omega$ the wind must start supersonic outside of the critical point in order to be supersonic at infinity.
%\begin{figure}[h!]
%\includegraphics[width=.99\columnwidth]{Parker-rot0.pdf} 
%\caption{Phase diagram for critical values $R_\Omega = r_0$.  The bifurcation point is at  $r=r_0/2$. For large $R_\Omega$ the wind must start supersonic.}
% \label{Parker-rot0}
%\end{figure}

\subsection{Connection to the base}
\begin{figure}[h!]
\includegraphics[width=.99\columnwidth]{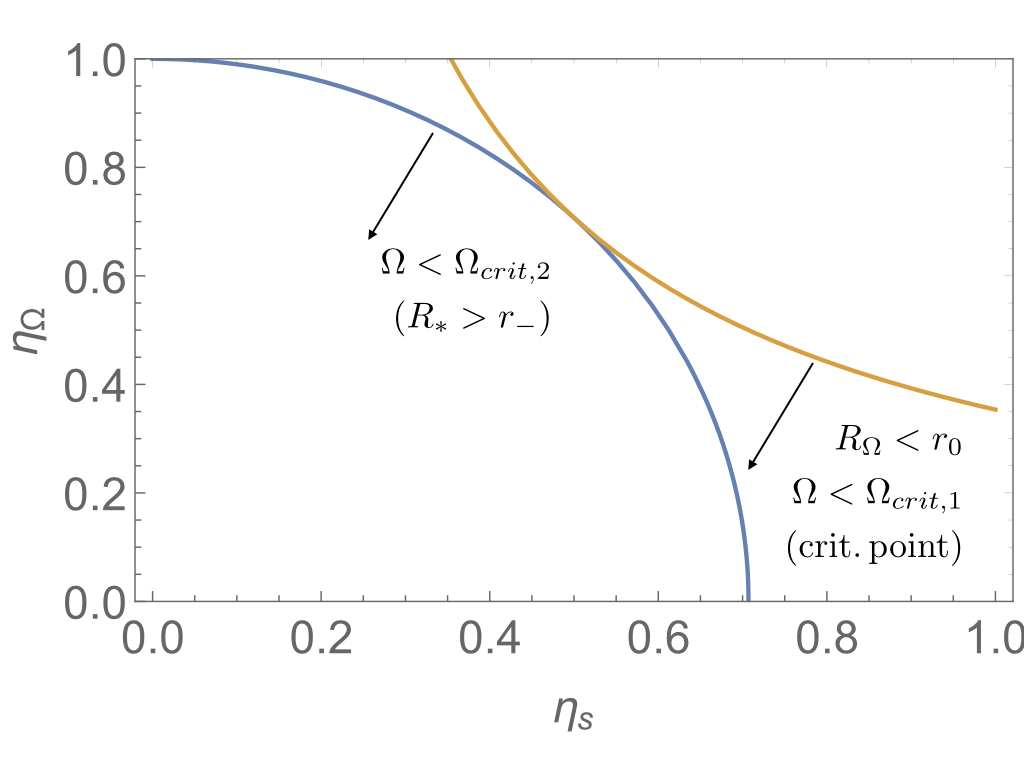} 
\caption{Allowed regions in the $\eta_s - \eta_\Omega$ plane.}
 \label{Parker-Roche-parameters}
\end{figure}

The radius of the star  $R_{b}$ cannot be smaller than $r_{min}$ for the model to be applicable.
If the  radius of the star  $R_{b}$ is  larger than $r_-$,   then the wind  continuously accelerate. 
The value of $r_{min}$ is very close to analytical $r_-$. The requirement $R_{b} \geq r_-$   translates to
 $\Omega <\Omega _{crit, 2}$, 
\ba &&
\Omega _{crit, 2} = \sqrt{ \Omega_K ^2 - 2 (c_s/R_{b})^2}
\nn &&
\frac{c_s}{ R \Omega_K} \leq \frac{1}{\sqrt{2}} \sqrt{ 1 - (\Omega/\Omega_K)^2}
\label{constr}
\ea
where   $\Omega_K= \sqrt{ G M_\ast /R_{b}^3}$ is Keplerian angular velocity at equator, see Fig. \ref{Parker-Roche-parameters}

There is also a special point where (\ref{const2}) and (\ref{constr}) match
\ba &&
\eta_\Omega =\frac{1}{\sqrt{2}}
\nn &&
\eta_s = \frac{1}{2 }
\nn &&
R_\Omega = r_0,\, \Omega_{max}= \Omega_K/2, \, r_+ = r_0/2
\ea
In this case the wind starts sonically right from the surface $R_{b}$ and evolves according to 
\be
-\frac{M_s^2}{2}  + \ln M_s =\frac{5}{2} - 4 \frac{R_{b}}{r} +  \left(\frac{R_{b}}{r} \right)^2 + 2 \ln \frac{R_{b}}{r} 
\ee

\subsection{Equatorial disk winds}

One of the possible applications of the model is for thermal winds launched by disks \citep[eg][]{WeberDavis,2000prpl.conf..759K,blandford_82,2018MNRAS.481.2628W}.  Assuming thin disk, so that the flow stars from a Keplerian-moving base, In this case then  $\Omega=\Omega_K$, $\eta_\Omega=1$,   $R_{b}$  means the local radius of the disk, and $c_s$ refers to the sound speed in the corona above the disk. Our parameters become
\ba &&
R_\Omega = \sqrt{ 2 G M_\ast R_{b}}/c_s = 2\sqrt{2} \eta_s r_0
\nn &&
\eta_\Omega=1
\nn &&
r_+ = \frac{1}{2} \left(1+\sqrt{1-8 \eta_s^2}\right) r_0
\ea

Thus, to have a critical point (to have a subsonic region) it is required that  $\eta_s< 1/(2\sqrt{2})=0.354$, or in physical units, 
\be
c_s \leq \frac{R_{b} \Omega_K}{2 \sqrt{2}} = \frac{v_K}{4},
\ee
For larger  coronal sound speed there is no critical point.  
 
 There is also the condition that $r_{min}$ should be smaller than $R_{b}$  for a flow to accelerate outwards.  Here we cannot  approximate $r_{min}$ as $\sim r_-$ (this would give only a trivial solution $\eta_s=0$). For the Keplerian disk  
 $R_{b}= R^2_\Omega/(4 r_0)$  and, using the calculation of $r_{min}$, Fig. \ref{rmin}, we find that in order to start subsonically the coronal sounds speed should satisfy
 \be
 \eta_s \leq 0.3190,
 \ee
 or, in physical units
 \be
 c_s < 0.3190 R_{b} \Omega_K= 0.225 v_K
 \ee
 (for larger  coronal sound speeds   the critical curve does not reach a given point on the disk).

 \section{Discussion}

 In this paper we consider a highly idealized problem of thermal wind launched from a rotating object, a star or  a disk.  Our analytical results seem to be in agreement with previous numerical works by 
 \cite{2010ApJS..188..290S,2012MNRAS.426.2239W}. In particular, \cite{2012MNRAS.426.2239W} argued for a single critical point and also found regimes of non-continuous accelerations, ``enthalpy deficit regime''.
 
 Our results differ from the case of radiation-driven rotating winds \citep{1986ApJ...311..701F}. In that case, \eg\ the critical point moves outward due to rotation, while the terminal velocity is smaller. In our case the  critical point moves inward, while at each radius the velocity is higher than in the non-rotating case. The critical conditions in line-driven winds are qualitatively different from the pressure-driven winds, \citep[\eg][]{1999isw..book.....L}.

 It is of interest to discuss the cases of high rotation rates/high sound speeds when the model breaks down. There are two constraints: (i) connection to the base, $r_{min} \geq R_b$; (ii) existence of critical points, $R_\Omega < r_0$. If the condition (i) is broken, then the only way for a subsonic flow to connect to infinity is through  unphysical ``breeze" solution (it is subsonic, but typically has much larger pressure that prevents a smooth match to the interstellar medium).
 Similarly, if $R_\Omega > r_0$ and the flow is subsonic at the base, the breeze solution is the only choice. These cases are somewhat different from the classical Parker model, where the critical subsonic curve extends to $M_s=0$ as $r \rightarrow 0$.  In our case all subsonic breeze solutions connect to supersonic non-critical solutions at $M_s=1$. In this regimes, most likely, the flow either  becomes non-stationary and/or may form shocks. 
 
 The generalization to the polytropic case should be straightforward and follow the classic Parker's case: instead of continuous acceleration, a  supersonic branch of the flow would reach  a constant velocity. 

\section*{Acknowledgments}
This work has been supported  by  DoE grant DE-SC0016369 and NASA grant 80NSSC17K0757.

I would like to thank Maxim Barkov,  Zhuo Chen, Daniel Proga and Tim Waters for discussions.

\bibliographystyle{apj}

 \bibliography{/Users/maxim/Home/Research/BibTex}
 
 \appendix
 
 \section{Comparing with  \cite{2010ApJS..188..290S}   - Parker wind in cylindrical geometry}
 \label{cyll}
 
 In  cylindrical geometry  Eq. (\ref{mdot}) changes to 

 \be
r \rho v_r={\rm Constant}
\label{mdotc}
\ee
while the  wind equation becomes
\be
\frac{v_r '}{v_r}  = \frac{  c_s^2 r^2 + r^3 \Phi' (r) + R_{b}^4 \Omega^2}{r^3(v_r^2 -c_s^2)}
\label{1c}
\ee
It differs from  the spherical case, Eq. (\ref{1}), only by a different factor ($1$ instead of $2$) in front of the $c_s^2$ term in the numerator.  The structure of the equation remains the same, only with slightly changed definitions of the parameter (\eg, the location of critical points remain the same as (\ref{rcrit}) but with $r_0 = G M/c_s^2$  and  $R_\Omega = 2 R_b^2 \Omega/c_s$).

 \end{document}